\documentstyle[12pt]{article}
\input epsf

\def\theequation{{\arabic{section}.\arabic{equation}}}
\def\reset{\setcounter{equation}{0}}

\def\pr#1#2#3#4{{\it Phys. Rev. D\/}{\bf #1}, #2 (19#3#4)}
\def\np#1#2#3#4{{\it Nucl. Phys.} {\bf B#1}, #2 (19#3#4)}

\begin{document}
\hsize=6.5truein
\hoffset=-.5truein

\begin{titlepage}
\null\vspace{-62pt}

\pagestyle{empty}
\begin{center}
\rightline{CU-TP-704}
\rightline{hep-ph/9507381}

\vspace{1.0truein}
{\Large\bf Gauge independence of the bubble  nucleation \\rate in
                  theories with radiative symmetry breaking}

\vspace{.2in}
Dimitrios Metaxas and Erick J. Weinberg\\

{\it Department of Physics, Columbia University,
New York, NY  10027}\\

\vspace{0.5in}

 \title{\Large\bf Gauge independence of the bubble nucleation rate in
                  theories with radiative symmetry breaking}
 \author{\large\rm Dimitrios Metaxas and Erick J. Weinberg\\ \\
  \large\it Department of Physics\\ \large\it Columbia University\\
 \large\it New York, NY 10027}

\vspace{.5in}
\centerline{Abstract}
\begin{quotation}
\baselineskip 24pt
    In field theories where a metastable false vacuum state arises as
a result of radiative corrections, the calculation of the rate of
false vacuum decay by bubble nucleation depends on the effective
potential and the other functions that appear in the derivative
expansion of the effective action.  Beginning with the Nielsen
identity, we derive a series of identities that govern the gauge
dependence of these functions.  Using these, we show, to leading
nontrivial order, that even though these functions are individually
gauge-dependent, one obtains a gauge-independent result for the bubble
nucleation rate.  Our formal arguments are complemented by explicit
calculations for scalar electrodynamics in a class of $R_\xi$ gauges.

\end{quotation}
\end{center}

\end{titlepage}

\newpage
\pagestyle{plain}
\setcounter{page}{1}
\newpage

\section{\rm Introduction}

    In addition to the minimum energy true vacuum state, many quantum
field theories have one or more metastable ``false vacua'' that can
decay to the true vacuum by the nucleation of bubbles of the stable
vacuum.  Methods have been developed for calculating the rate of this
process either at zero temperature \cite{Coleman} or at high
temperature \cite{Linde}.  However, these must be modified in the case
of theories in which symmetry breaking arises as a result of radiative
corrections \cite{CW}.  While a scheme for dealing with such cases (at
zero temperature) has been developed \cite{EJW}, it leads to an
expression for the bubble nucleation rate that is not manifestly
gauge-independent.  In this paper we address the issue of this gauge
dependence.

     The standard approach \cite{Coleman} to the calculation of the
bubble nucleation rate at zero temperature is based on finding a
``bounce'' solution of the classical Euclidean field equations.  The
nucleation rate per unit volume $\Gamma$ may be written as
\begin{equation}
     \Gamma = A e^{-B}
\end{equation}
where B is the Euclidean action of the bounce solution and A is an
expression involving functional determinants that is generally equal
to a numerical factor of order unity times a dimensionful quantity
determined by the characteristic mass scales of the theory.

   A problem arises if radiative corrections modify the vacuum
structure of the theory.  Theories in which this happens
generally have no bounce solution; even if a bounce does exist,
the nucleation rate calculation based on the bounce is not
reliable.  However, by integrating out certain fields at the
outset, one can derive a modified algorithm \cite{EJW} that can be applied
to this situation.  The results of this method are conveniently
expressed in terms of the functions that appear in the
derivative expansion
\begin{equation}
     S^{\rm eff} = \int d^4x \left[V^{\rm eff}(\phi) + {1\over 2} Z(\phi)
      (\partial_\mu \phi)^2 + \cdots \right]
     \label{derivexpansion}
\end{equation}
of the Euclidean effective action.  (The dots represent terms containing
four or more derivatives; these do not enter the calculation at the order
to which we work.) These
functions can in turn be expanded in power series in the couplings.
For example, in a gauge theory with weak
(i.e., $O(e^4)$) scalar self-couplings, the effective potential is of order
$e^4$ and may be written, using an obvious notation, as
\begin{equation}
     V^{\rm eff} = V^{\rm eff}_{e^4} + V^{\rm eff}_{e^6} + \cdots \, ,
     \label{Veffexpansion}
\end{equation}
while
\begin{equation}
     Z = 1 + Z_{e^2} + \cdots \, .
\end{equation}

    The first step in this approach is to use the leading approximation
to the effective action to
determine a bounce solution $\phi_b(x)$ through the equation
\begin{equation}
     \Box\phi_b =\frac{\partial V_{e^4}^{\rm eff}}{\partial\phi} \, .
    \label{bounceeq}
\end{equation}
The desired nucleation rate is then given by
\begin{equation}
      \Gamma =A^\prime e^{-(B_0 +B_1)}
\end{equation}
where
\begin{equation}
   B_0 = \int d^4x \left[V_{e^4}^{\rm eff}(\phi_b) +\frac{1}{2}
  (\partial_\mu \phi_b)^2 \right]
\end{equation}
turns out to be of order $e^{-4}$ while
\begin{equation}
    B_1 = \int d^4x \left[V_{e^6}^{\rm eff}(\phi_b) +
   \frac{1}{2} Z_{e^2} (\phi_b) (\partial_\mu \phi_b )^2 \right]
     \label{Bonedef}
\end{equation}
is of order $e^{-2}$.  The calculation of the
pre-exponential factor is much more complicated than in the standard
case; in particular $A'$ cannot be expressed solely in terms
of the functions appearing in Eq.~(\ref{derivexpansion}).
Nevertheless, one finds that, just as in the standard case,
$A'$ is equal to a numerical factor of order unity times a
dimensionful factor determined by the mass scales of the theory.

   Like any physically measurable quantity, the nucleation rate should
be gauge independent.  Since the leading terms in the effective
potential are gauge independent, there is no difficulty in this regard
with respect to either $B_0$ or the bounce solution itself.  However,
both of the functions that enter in $B_1$ are known \cite{Jackiw} to
depend on gauge.  Our goal is to show that, nevertheless, these
combine to give a gauge-independent contribution to the nucleation rate.
Although we do not explicitly examine the prefactor $A'$, we expect
that our methods could be extended --- albeit with considerably more
technical complication --- to show that it too is independent of gauge.

     Our approach is based on the Nielsen identity \cite{Nielsen},
which describes the gauge dependence of the effective action, and
which has been used to show that gauge-independent physical quantities
can be obtained from a gauge-dependent effective potential
\cite{Fukuda}. In
Sec.~2 we present a compact derivation of this identity, following the
method of Kobes, Kunstatter, and Rebhan \cite{Kobes}.  However, the
usual form of the identity is not quite sufficient for our purposes.
Instead, what we need is a series of identities, each of which gives
the gauge dependence of one of the functions appearing in the
derivative expansion (\ref{derivexpansion}).  Although the identity
for the effective potential is well-known, the remaining identities
are, to our knowledge, new.  In Sec.~3 we derive these from the master
identity and then use them to give a general proof of the
gauge-independence of $B_1$.  To complement this formal proof, we have
verified the relevant identities by explicit calculations for the case
of scalar quantum electrodynamics in $R_\xi$ gauges.  These
calculations, which expand upon the work of Aitchison and Fraser
\cite{AandF}, are described in Sec.~4.  Section~5 contains some
concluding comments.  Some two-loop effective potential calculations
are presented in an Appendix.

\section{The Nielsen identity}
\reset

   In this section we use the method of Ref.~\cite{Kobes} to derive
the Nielsen identity.  We consider a gauge theory with fields denoted
by $\phi_i$.  The classical action $S$ is invariant under a set of
infinitesimal gauge transformations of the form
\begin{equation}
    \delta_{g}\phi_{i}=\Delta_i^\beta \theta_\beta
       \label{gaugetrans}
\end{equation}
where the $\Delta_i^\beta$ are linear operators.  ( We will henceforth
suppress the index $\beta$; for scalar electrodynamics, which we
examine in greatest detail, there is only a single gauge parameter
$\theta$ in any case.)  By choosing a gauge-fixing function
$F(\phi_i)$ and introducing
Fadeev-Popov ghosts $\eta$ and $\bar\eta$, we can write the
generating functional of connected Green's functions as
\begin{equation}
	 \exp\left[iW(J,F)\right]=\int[{\cal D}\phi_i] [{\cal D}\eta]
     [{\cal D}\bar{\eta}]
         \exp\left[iI(F) +i \int d^4y J^i(y)\,\phi_i(y)\right]
     \label{Wdef}
\end{equation}
where
\begin{equation}    I(F)=S - \int d^4x
      \left\{\frac{\left[ F(\phi)\right]^{2}}{2\xi} +
  \bar{\eta}\frac{\delta F}{\delta\phi_{i}}\Delta_{i}\eta   \right\}
   \label{Idef}
\end{equation}
is invariant under the BRST transformations
\begin{equation}    \delta_{{\rm BRST}}\,\phi_{i}=\zeta \,\Delta_{i}\eta,
\qquad
    \delta_{{\rm BRST}}\,\bar\eta= -\zeta  \frac{1}{\xi} F, \qquad
    \delta_{{\rm BRST}}\,\eta =0
         \label{BRST}
\end{equation}
with $\zeta $ an arbitrary Grassman number.

    In presenting the derivation, it is convenient to adopt a
compact notation where
\begin{equation}
   \left\langle{\cal O}(\phi)\right\rangle\equiv
        e^{-iW}\int[{\cal D}\phi] [{\cal D}\eta] [{\cal D}\bar\eta]
     {\cal O}(\phi,\eta, \bar\eta)
      \exp \left[i I + i \int d^4y J^i(y)\phi_i(y) \right]
\end{equation}
for any operator $\cal O$.  Now note that if $\cal O$ is
linear in the ghost fields, its odd Grassman character leads
to the vanishing of this quantity.  In particular,
\begin{equation}
      \langle\bar\eta \,  G \rangle=0
\end{equation}
for any functional $G[\phi(x)]$.  Applying the BRST transformation
(\ref{BRST}) to this equation results in the identity
\begin{equation}
   \left\langle\delta_{{\rm BRST}}[\bar\eta(x)\, G(x) ]
  +i \bar\eta(x) G(x) \int d^4 y J^i(y)\, \delta_{{\rm BRST}}\phi_i(y)
      \right\rangle=0
\end{equation}
which may be rewritten, using the anticommutivity of
$\eta$ and $\bar\eta$, as
\begin{equation}
  -\left\langle \frac{1}{\xi} F(x)\, G(x) +
   \bar\eta(x) \frac{\delta G(x)}{\delta\phi_i(x)} \,\Delta_i\eta(x)
      \right\rangle = -i \int d^4y J^i(y) \left\langle
        \Delta_i \eta(y) \, \bar\eta(x)\, G(x)  \right\rangle \, .
    \label{FGeq}
\end{equation}

     Now consider the effect of an infinitesimal change $F \to F +
\Delta F$ in the gauge-fixing function.  Recalling Eqs.~(\ref{Wdef})
and (\ref{Idef}),
we see that the change in the generating functional $W[J]$ is simply
the integral over $x$ of the left hand side of Eq.~(\ref{FGeq}), with
$G$ set equal to $\Delta F$.  Hence,
\begin{equation}
   \Delta W= - i \int d^4x\, d^4y \, J^i(y) \langle\Delta
_i\eta(y)\,\bar\eta(x) \,
   \Delta F(x) \rangle    \, .
\end{equation}
Recalling that the effective action is related to $W[J]$ by the
Legendre transformation
\begin{equation}
   S_{\rm eff} = W - \int d^4x J^i(x) \, {\delta W \over\delta
   J_i(x)} = W - \int d^4x J^i(x)\, \phi_i(x)
\end{equation}
we find that
\begin{equation}
     \Delta S_{\rm eff} = i \int d^4x\, d^4y\,
    \frac{\delta S_{\rm eff}}{\delta\phi_i(y)}
      \langle \Delta_i\eta(y)\, \bar\eta(x)\, \Delta F(x)
          \rangle_{\rm 1PI}
    \label{DeltaSequation}
\end{equation}
where the subscript $\rm 1PI$ indicates that only the
contributions from one-particle irreducible graphs are to be
included. In particular, an infinitesimal change $d\xi$ in the
gauge parameter is equivalent to the choice $\Delta F = -(F/2
\xi) d\xi$.  Hence,
\begin{eqnarray}
  &&   \xi \frac{\partial S_{\rm eff}}{\partial\xi}
   = - \frac{i}{2} \int d^4x \,d^4y
    \frac{\delta S_{\rm eff}}{\delta\phi_i(y)}
    \langle \Delta_i\eta(y)\,\bar\eta(x)\,  F(x) \rangle_{\rm 1PI} \\
  &&\qquad\quad            = \int d^4y
         \frac{\delta S_{\rm eff}}{\delta\phi_j(y)} H_j[\phi(z),y]
     \label{basicID}
\end{eqnarray}
where
\begin{equation}
     H_j[\phi(z),y]  = -\frac{i}{2} \int d^4x
     \langle \Delta_i\eta(y)\,\bar\eta(x)\,  F(x) \rangle_{\rm 1PI} \,\, .
\end{equation}
Eq.~(\ref{basicID}) is the Nielsen identity.

\section{Derivative expansion of the Nielsen identity and formal
proof of gauge independence of the nucleation rate}
\reset

      To study the gauge-dependence of the bubble nucleation rate, we
will need a set of identities that are obtained by making derivative
expansions of both sides of Eq.~(\ref{basicID}).  For
simplicity, consider the case where the effective action depends on
only a single field $\phi(x)$.  There is then only a single functional
$H[\phi(x), y]$, which can be expanded as
\begin{equation}
       H[\phi(x),y] = C(\phi) + D(\phi)(\partial_{\mu}\phi)^2 +\cdots
\end{equation}
where all terms on the right are understood to be evaluated at point
$y$ and the dots represent terms with more than two derivatives.
Inserting this, together with the expansion (\ref{derivexpansion}) of
the effective action, into Eq.~(\ref{basicID}), gives
\begin{eqnarray}
    &&   \xi \frac{\partial}{\partial\xi}
      \int d^4x \left[V^{\rm eff}(\phi)
     + \frac{1}{2} Z(\phi)(\partial_\mu\phi)^2
     + \cdots \right]  \nonumber \\
   &&\qquad\quad =  \int d^4x \Bigg[C(\phi) + D(\phi)(\partial_\mu\phi)^2
       +\cdots \Bigg]
      \left[ \frac{\partial V^{\rm eff}}{\partial\phi} +\frac{1}{2}
      \frac{\partial Z}{\partial\phi} (\partial_\mu\phi)^2
   - \partial_\mu \left[ Z(\phi)\,\partial_\mu \phi \right] + \cdots
\right] \, .
    \nonumber \\
\end{eqnarray}

    If this identity is to hold for arbitrary $\phi(x)$, then not only
must the integrands on the two sides be equal point by point, but the
terms with equal number of derivatives must be separately equal.
Thus, the terms with no derivatives obey
\begin{equation}
    \xi \frac{\partial V^{\rm eff}}{\partial\xi} =
     C \frac{\partial V^{\rm eff}}{\partial\phi}
    \label{Videntity}
\end{equation}
while from the terms with two derivatives we obtain
\begin{equation}
    \xi \frac{\partial Z}{\partial\xi} =
           C \frac{\partial Z}{\partial\phi}
         + 2 D\frac{\partial V^{\rm eff}}{\partial\phi}
         + 2 Z\frac{\partial C}{\partial\phi} \, .
    \label{Zidentity}
\end{equation}
(Eq.~(\ref{Videntity}), which can be obtained immediately from
Eq.~(\ref{basicID}) by choosing $\phi(x)$ to be a constant,
appears in Ref.~\cite{Nielsen}.)

     We now specialize to the case of
a gauge theory with gauge coupling $e$ and scalar self-couplings of
order $e^4$.  As indicated in Eq.~(\ref{Veffexpansion}), the effective
potential begins
with terms of order $e^4$, while $Z(\phi) = 1 + O(e^2)$.  Analysis of the
relevant graphs shows that $C(\phi)$ starts at order $e^2$ and
$D(\phi)$ is of order unity.  The terms of order $e^4$
in Eq.~(\ref{Videntity}) yield
\begin{equation}
   \xi \frac{\partial V^{\rm eff}_{e^4}}{\partial\xi} =0  \, .
   \label{Ve4identity}
\end{equation}
Now recall that the bounce solution $\phi_b(x)$ is determined,
through Eq.~(\ref{bounceeq}), by $V^{\rm eff}_{e^4}$.  Since
Eq.~(\ref{Ve4identity}) shows that the latter is
gauge independent, both $\phi_b(x)$ and $B_0$, the leading
contribution to the exponent of the nucleation, are independent
of $\xi$.

    To study the gauge dependence of $B_1$, we need the order
$e^6$ terms of Eq.~(\ref{Videntity}),
\begin{equation}
   \xi \frac{\partial V^{\rm eff}_{e^6}} {\partial\xi}
    =  C_{e^2}\,  \frac{\partial V^{\rm eff}_{e^4}}{\partial\phi}   \, ,
   \label{Ve6identity}
\end{equation}
as well as the terms of order $e^2$ in
Eq.~(\ref{Zidentity}),
\begin{equation}
   \frac{1}{2} \xi\frac{\partial Z_{e^2}} {\partial\xi}
     =   \frac{\partial C_{e^2}}{\partial\phi} \, .
   \label{Ze2identity}
\end{equation}
These equations, together with Eq.~(\ref{Bonedef}), imply that
\begin{eqnarray}
      \xi \frac{\partial B_1}{\partial\xi} &=&
        \xi \frac{\partial}{\partial\xi} \int d^4x \left[
      V^{\rm eff}_{e^6} +
      \frac12 Z_{e^2} (\partial_\mu\phi_b)^2  \right]  \nonumber \\
     &=& \int d^4x \left[
      C_{e^2} \frac{\partial V^{\rm eff}_{e^4}}{\partial \phi}
      + \frac{\partial C_{e^2}}{\partial \phi} (\partial_\mu\phi_b)^2
       \right]   \nonumber \\
     &=& \int d^4x \left[
      C_{e^2} \frac{\partial V^{\rm eff}_{e^4}}{\partial \phi}
        + (\partial_\mu C_{e^2}) (\partial_\mu \phi_b) \right] \nonumber \\
     &=& \int d^4x \, C_{e^2}
    \left[ \frac{\partial V^{\rm eff}_{e^4}}{\partial \phi} - \Box \phi_b
        \right]
\end{eqnarray}
where all quantities are to be evaluated with $\phi(x)$ set equal to
the bounce solution $\phi_b(x)$.  Eq.~(\ref{bounceeq}), which
determines the bounce, shows that the last expression on the right
hand side must vanish, and hence that
\begin{equation}
      \xi \frac{\partial B_1}{\partial\xi} =0 \, .
\end{equation}
This verifies that, at least up to pre-exponential terms of
order unity, the bubble nucleation rate is gauge independent.

\section{Scalar Electrodynamics}
\reset
\subsection{Basics}

     We now illustrate these formal arguments by explicit
calculations for the case of scalar electrodynamics.
The Lagrangian, which we write in the form
\begin{equation}
  {\cal L}=-\frac{1}{4}F_{\mu\nu}^2
    +\frac{1}{2} (\partial_\mu\Phi_1 - e A_\mu \Phi_2)^2+
     \frac{1}{2} (\partial_\mu \Phi_2 + e A_\mu \Phi_1)^2
       - V(\Phi)
        \label{QEDLag}
\end{equation}
with
\begin{equation}
      V(\Phi) =
      \frac{1}{2}m^2 \Phi^2
      + \frac{\lambda}{4!} \Phi^4
\end{equation}
and $\Phi \equiv (\Phi_1^2 +\Phi_2^2)^{1/2}$,
is invariant under the gauge transformation
\begin{equation}
     \delta_g A_\mu = \partial_\mu \theta,
     \qquad \delta_g \Phi_1 = e \Phi_2 \theta,
      \qquad \delta_g \Phi_2 = - e \Phi_1 \theta  \, .
        \label{QEDgauge}
\end{equation}
If $m^2 >0$, the tree-level potential has a minimum at $\Phi=0$.
In order that one-loop effects be able to change the vacuum
structure and give a symmetry-breaking minimum at $\Phi =
\langle \Phi\rangle \ne 0$, we must require that both $\lambda$
and $m^2$ be anomalously small, of order $e^4$ and $e^2 \langle
\Phi\rangle^2$, respectively.

     For calculating the bubble nucleation rate it is sufficient to
evaluate the terms in the derivative expansion of the effective action
for $\Phi_2=A_\mu=0$.  With this in mind, we will consider the class
of gauges determined by the gauge-fixing function
\begin{equation}
    F= \left( \partial_\mu A^\mu + ev \Phi_2\right)  \, .
    \label{gaugefunction}
\end{equation}
(The gauge-dependence of the effective potential in these gauges was
studied in detail by Aitchison and Fraser \cite{AandF}; in the following
discussion we will make use of a number of their results.)  The
Nielsen identity (\ref{basicID}) then involves only the single
functional
\begin{equation}
     H_{\Phi_1}[\phi(z),y]  = -\frac{ie}{2} \int d^4x
     \langle (\Phi_2(y)\,\eta(y)\,\bar\eta(x)\,
    \left( \partial_\mu A^\mu(x) + ev \Phi_2(x)\right) \rangle \, .
    \label{Hforxi}
\end{equation}

    The effective action can be obtained as the sum of one-particle
irreducible vacuum graphs in the theory obtained from the Lagrangian
(\ref{QEDLag}) by making the shift $\Phi_1 \to \Phi_1 + \phi $ and
then dropping all terms linear in the quantum fields.  The vertex
factors for these graphs can be simply read off from the resulting
Lagrangian in the standard fashion (see, e.g., Ref.~\cite{AandF}).
The propagators require a bit more work.  Following the usual
approach, one would obtain from the Lagrangian (together with the
gauge-fixing and ghost terms) the effective $\Phi_1$, $\Phi_2$,
$A_\mu$, and ghost propagators
\begin{eqnarray}
&&    G_1(k)= {i \over k^2 - {\tilde m}_1(\phi)^2 } \label{phi1prop}\\
&&    G_2(k) = {i(k^2 -\xi e^2\phi^2) \over  D(k)}  \label{phi2prop}\\
&&    G_{\mu\nu}(k) = G^T_{\mu\nu}(k) +  G^L_{\mu\nu}(k)
          \nonumber \\
&&  \qquad      = i\frac{-g_{\mu\nu} + \frac{ k_\mu k_\nu}{k^2}}
{k^2-e^2\phi^2}
       - \frac{i [ \xi (k^2 -{\tilde m}^2_2) -e^2v^2 ]}{D(k)}
       \frac{ k_\mu k_\nu}{k^2 } \label{photonprop} \\
&&    G_g = {i \over k^2 + e^2v\phi } \label{ghostprop}
\end{eqnarray}
as well as the mixed $\Phi_2\,$--$A_\mu$ propagator
\begin{equation}
     G_{2\mu}(k) =  { e (\xi \phi + v)\, k_\mu \over D(k) }
\end{equation}
where the momentum flow is understood to flow from the $\Phi_2$
end to the $A_\mu$ end.  In these expressions
\begin{equation}
      D(k) = k^4 - k^2 ( {\tilde m}^2_2  -2e^2 v\phi)
    + e^2 \phi^2 (e^2v^2 + \xi {\tilde m}_2^2 )
    \label{Ddef}
\end{equation}
and
\begin{eqnarray}
   &&  {\tilde m}^2_1(\phi) = m^2 + \frac\lambda{2} \phi^2
  =  V''(\phi)  \\
   &&   {\tilde m}^2_2(\phi)=  m^2 + \frac\lambda{6} \phi^2
   =  \frac{V'(\phi)}{\phi}  \, .
\end{eqnarray}

    These propagators are not quite what we need.  Our assumption that
$\lambda$ is of order $e^4$ not only makes some of the one-loop terms
comparable to the tree-level terms, but also implies that some
multi-loop graphs are not suppressed relative to graphs with fewer
loops; specifically, the insertion of transverse photon loops along a
scalar propagator does not increase the order of the graph.  To
restore the validity of our expansion, these insertions must be
summed.  This can be done simply by replacing the propagators given in
Eqs.~(\ref{phi1prop}-\ref{photonprop}) by ``dressed'' propagators in
which the ${\tilde m}_a^2$ are replaced by
\begin{eqnarray}
   &&  m^2_1(\phi) =  V''_{e^4}(\phi)  \\
   &&  m^2_2(\phi) = \frac{V'_{e^4}(\phi)}{\phi} \, .
   \label{m2mass}
\end{eqnarray}
(To avoid double-counting, subtractions are needed for certain graphs
with two or more loops; these corrections only affect contributions of
higher order than those we will be considering.)

    Before we proceed to verify the identities, there is one more
issue to be addressed.  Many of the graphs contributing to the
effective action have divergences that must be cancelled by
appropriate counterterms.  We will not display these explicitly, but
all divergent integrals should be understood to be made finite by some
gauge-invariant renormalization scheme (e.g., minimal subtraction in
the context of dimensional regularization); when we refer to the
magnitude of an integral, this should be understood as referring to
the magnitude of its finite part.

\subsection{The identity for the effective potential}

     To order $e^2$ the function $C(\phi)$ entering the identity
(\ref{Videntity}) receives contributions only from the two
graphs shown in Fig.~1.  These combine to give \footnote[1]{~Apart
from an overall sign arising from a difference in the definition in
$C(\phi)$, these expressions are the same as those appearing in
Ref.~\cite{AandF}}
\begin{eqnarray}
&&     C_{e^2} = -{ie\over 2} \int{ d^4k \over (2\pi)^4}
      {1 \over (k^2 + e^2 v \phi)\, D(k)}
   \left[ e (\xi \phi +v)\, k^2   - ev( k^2 -\xi e^2 \phi^2 ) \right]
            \nonumber \\
&&  \quad \,\,\,\,= -{ie^2 \phi \xi\over 2}
         \int{ d^4k \over (2\pi)^4} {1\over D(k)} \, .
    \label{Ce2}
\end{eqnarray}

     The effective potential is obtained by summing the graphs with
vanishing external momenta.  The one-loop contributions may be split
into three parts.  First, the graphs with a transverse photon loop
give a contribution
\begin{equation}
       - \frac{3i}{2} \int{ d^4k \over (2\pi)^4}
   \ln (k^2-e^2\phi^2) \, .
\end{equation}
{}From dimensional arguments, this integral is clearly of order $e^4
\phi^4$.  Hence, it combines with the tree-level potential to give
\begin{equation}
     V_{e^4}^{\rm eff} = \frac{1}{2} m^2 \phi^2 + \frac{\lambda}{4!}\phi^4
     -\frac{3i}{2} \int {d^4k\over (2\pi)^4} \ln (k^2-e^2\phi^2) \, .
\end{equation}
This is manifestly gauge-independent, in accordance with
Eq.~(\ref{Ve4identity}).
A second gauge-independent contribution, coming from the graphs
with $\Phi_1$-loop graphs, is
\begin{equation}
       -\frac{i}{2} \int {d^4k\over (2\pi)^4} \ln (k^2 - m_1^2 ) \, .
\end{equation}
This is of order $m_1^4$, and hence contributes only to $V_{e^8}^{\rm eff}$.
Finally, there is an order $e^6$ contribution
\begin{equation}
       V_{e^6; {\rm 1-loop}}^{\rm eff} =
          -\frac{i}{2} \int{ d^4k\over (2\pi)^4} \left [\ln D(k) -
     2\ln (k^2 + e^2 v \phi) \right]
    \label{Ve6oneloop}
\end{equation}
in which the first term arises from graphs with $\Phi_2$, longitudinal
photon, or mixed scalar-photon propagators while the second is due to
those with a single ghost loop.

     In addition to these one-loop contributions, there are a number
of two-loop graphs that contribute to $V^{\rm eff}_{e^6}$.  Some
two-loop graphs have already been included in Eq.~(\ref{Ve6oneloop})
as a result of the replacement ${\tilde m}^2_i \to m^2_i$, including
in particular the $\xi$-dependent ``figure-eight'' graph with one
transverse photon loop and one $\Phi_2$ loop.  In the Appendix we show
that although a number of the remaining graphs are $\xi$-dependent,
they add together\footnote[2]{~The cancellation of the gauge-dependence
among these graphs can be understood by considering the case $\lambda
\sim e^2$, where the loop expansion is completely equivalent to an
expansion in $e^2$.  Apart from the appearance of ${\tilde m}^2_2$
rather than $m_2^2$, the one-loop approximation to $C(\phi)$ is
precisely the same as $C_{e^2}$ of Eq.~(\ref{Ce2}).  Hence,
Eq.~(\ref{Videntity}) can be satisfied both at the one-loop level in
that case (as was shown in ~\cite{AandF}) and in at $O(e^6)$ in our
case only if this cancellation among the two-loop graphs occurs.}
to give a gauge-independent contribution to $V^{\rm eff}_{e^6}$.
Hence, we can combine Eq.~(\ref{Ve6oneloop}) with
Eqs.~(\ref{Ddef}), (\ref{m2mass}), and (\ref{Ce2}), to obtain
\begin{equation}
   \xi \frac{\partial V^{\rm eff}_{e^6}} {\partial\xi} =
   \xi \frac{\partial V^{\rm eff}_{e^6; {\rm 1-loop}}} {\partial\xi}
    =  C_{e^2}\,  \frac{\partial V^{\rm eff}_{e^4}}{\partial\phi} \, ,
\end{equation}
thus verifying Eq.~(\ref{Ve6identity}).

\subsection{The identity for $Z(\phi)$}

    We now turn to the identity (\ref{Zidentity}), which we will
verify to order $e^2$.  We begin by recalling that
$Z(\phi)$ can be calculated from the sum of one-particle irreducible
graphs with one external line carrying momentum $p$, another carrying
momentum $-p$, and all others with zero momentum.  If the
contribution of graph $j$ is denoted by $I_j(p^2)$, then
\begin{equation}
       Z = - i \left.{\partial \over \partial p^2} \sum I_j
        \right|_{p^2=0}   \, .
\end{equation}

    Although there are many one-loop graphs contributing to the scalar
self-energy, we will need to calculate only a few.  Those graphs with
quartic vertices are independent of the external momentum and hence do
not contribute to $Z(\phi)$.  Because of our assumption that $\lambda
= O(e^4)$, all graphs with a vertex arising from the scalar
self-interaction are at least of order $e^4$ and can also be ignored
here.  Finally, the self-energy graph with a single ghost loop,
although of order $e^2$, is $\xi$-independent.  Thus, the entire
$\xi$-dependence of $Z_{e^2}(\phi)$ comes from the four graphs shown
in Fig.~2.

      It is convenient to consider separately the terms containing the
transverse and the longitudinal parts of the photon propagators.  The
only contribution with two transverse propagators, from
graph b, is manifestly independent of $\xi$ and so can be
neglected here.  Graphs a, b, and c each give contributions with a
single transverse propagator; although separately these each contain
$\xi$-dependent terms of order $e^2$, their sum is easily seen to be
$\xi$-independent to this order.

   This leaves the terms containing only longitudinal photons.  These
may be written as
\begin{equation}
     I_j = e^2 \int{d^4k \over (2\pi)^4}
     {1 \over D(k) \, D(p+k)} \, b_j(k,p)
   \label{graphintegral}
\end{equation}
where
\begin{eqnarray}
&&  b_a = - [(p+k)^2 -\xi e^2 \phi^2]\, (\xi k^2 -e^2v^2) \,
   [k \cdot (2p+k)]^2 \,k^{-2}  \nonumber \\ \nonumber \\
&& b_b = 2 e^2 \phi^2 \,(\xi k^2 -e^2v^2) \, [\xi (p+k)^2 - e^2v^2]
   [k \cdot (p+k)]^2 \,k^{-2} \,(p+k)^{-2} \nonumber \\ \nonumber   \\
&& b_c = -4 e^2\phi^2 \,(\xi \phi +v)\, (\xi k^2 -e^2v^2)
     [k\cdot (p+k)] \, [k\cdot (2p+k)] \, k^{-2} \nonumber \\ \nonumber\\
&& b_d = e^2 (\xi \phi +v)^2\, (k^2 -p^2) \, k\cdot(k+2p)  \, .
\end{eqnarray}
(In these expressions we have omitted terms proportional to $m^2_2$
since the contribution of these is at least $O(e^4)$.  Note also that
we have included a factor of two in the the contribution from graph c
to take into account the fact that reversing the direction of $p$
gives a second graph with the same value.)  Summing these expressions,
we obtain
\begin{eqnarray}
&&    b_a +b_b +b_c +b_d = \xi (p^2 - 2 p\cdot k - k^2) D(p+k)
      -2 \xi p^2 \left(e^2 \phi v + e^4 \phi^2 v^2 \right)
     \nonumber \\
&& \qquad \qquad
    +  \left\{ {e^6 \phi^4 v^2\over 2} \left[3(k+p)^2 - 3k^2
    + {(p+k)^4 \over k^2} - {k^4 \over (p+k)^2}  \right]
     + {e^4\phi^2\xi^2\over 2} [ k^4 - (p+k)^4 ] \right\}
   \nonumber\\ && \qquad  \qquad + \cdots
    \label{bjsum}
\end{eqnarray}
where the dots denote terms that are either $\xi$-independent, of
order $p^3$, or else proportional to $m_2^2$ and thus of higher order
in $e$.  Because of their antisymmetry under the interchange $k^2
\longleftrightarrow (p+k)^2$, the contributions of the terms in curly
brackets to the integral in Eq.~(\ref{graphintegral}) cancel.
Inserting the remaining terms into the integral and keeping terms
proportional to $p^2$, we find that
\begin{eqnarray}
&&  Z_{e^2} = -i e^2 \xi \int{d^4k \over (2\pi)^4} \left[ {1\over D(k)}
   - {2\left(e^2 \phi v + e^4 \phi^2 v^2 \right)\over [D(k)]^2}  \right]
     + \xi{\rm -independent\ terms}
 \nonumber \\  && \quad \, \, \,
 = -i e^2 \xi \int{d^4k \over (2\pi)^4} \left[ {1\over D(k)}
   - {\phi\over [D(k)]^2}  {\partial D(k) \over \partial \phi} \right]
       + \xi{\rm -independent\  terms}   \nonumber \\
&& \quad \,\,\,  = 2 {\partial C_{e^2} \over \partial \phi}
        + \xi{\rm -independent\ terms} \, .
\end{eqnarray}
(In going from the first to the second line, contributions proportional to
$\partial m_2^2/\partial \phi$ have been neglected as being of
higher order.)

    Differentiating this with respect to the
gauge-parameter $\xi$ gives
\begin{eqnarray}
&&    \xi {\partial Z_{e^2} \over \partial \xi} = 2 {\partial \over
   \partial \phi} \left[ C_{e^2} +ie^2 \phi \xi
    \int{d^4k \over (2\pi)^4}  {e^2 \phi^2 \xi m_2^2 \over [D(k)]^2}
      \right]  \nonumber \\
&& \qquad \,\,\,\,\, = 2 {\partial  C_{e^2} \over  \partial \phi} + O(e^4) \, .
\end{eqnarray}
This verifies the identity (\ref{Ze2identity}).

\section{Concluding remarks}
\reset

      In this paper we have have shown how the Nielsen identity that
describes the gauge dependence of the effective action can be
converted into an infinite series of identities, one for each of the
functions appearing in the derivative expansion of the effective
action.  Using these identities, we have shown, to leading nontrivial
order, that one obtains a gauge-independent result for the bubble
nucleation rate even in theories where the calculation of this rate
involves the gauge-dependent higher order contributions to the
effective action.  This provides one more example to show that the
gauge-dependence of the effective action does not prevent it from
being a useful tool for obtaining gauge-independent physical results.

     As an explicit example, we have verified the identities for the
$\xi$-dependence of $V^{\rm eff}(\phi)$ and $Z(\phi)$ in the class of
gauges defined by the gauge-fixing function (\ref{gaugefunction}).  In
fact, these gauges actually depend on a second parameter, $v$.
(Note that nothing in our calculations requires that $v$ be equal to
the vacuum expectation value of $\phi$.)  Working from
Eq.~(\ref{DeltaSequation}), we find that
\begin{equation}
    v \frac{\partial V^{\rm eff}}{\partial v} =
     C^v \frac{\partial V^{\rm eff}}{\partial\phi}
\end{equation}
and
\begin{equation}
    v \frac{\partial Z}{\partial v} =
           C^v \frac{\partial Z}{\partial\phi}
         + 2 D^v\frac{\partial V^{\rm eff}}{\partial\phi}
         + 2 Z\frac{\partial C^v}{\partial\phi} \, .
\end{equation}
where $C^v(\phi)$ and $D^v(\phi)$ are obtained from the derivative
expansion of
\begin{equation}
     H_{\Phi_1}^v[\phi(z),y]  = -ie^2 v \, \int d^4x
     \langle (\Phi_2(y)\,\eta(y)\,\bar\eta(x)\,
      \Phi_2(x) \rangle \, .
\end{equation}
(Note the absence of the factor of $1/2$ relative to
Eq.~(\ref{Hforxi}).)  In particular, the leading contributions to
these identities comes from
\begin{equation}
     C_{e^2} = - ie^2 v \int{ d^4k \over (2\pi)^4}
      {( k^2 -\xi e^2 \phi^2 ) \ \over (k^2 + e^2 v \phi)\, D(k)}
\end{equation}

     The Fermi gauges, defined by $F= \partial_\mu A^\mu$, can be
obtained from the $R_\xi$ gauges we have considered by setting $v=0$.
However, because of the infrared divergences that afflict these
gauges, the limit $v \to 0$ is somewhat nontrivial and the
verification of the identities for these gauges must be done
separately \cite{Stathakis}.  To see the cause of these difficulties,
note that our assumptions about the magnitude of $\lambda$ and $m^2$
imply that if $v=0$ the zeros of $D(k)$ occur at values of $k^2$ of
order $e^4 \phi^2$.  This has the effect of making some two-loop
graphs (beyond those resummed by the conversion of ${\tilde m}^2_i$ to
$m_i^2$) comparable to one-loop graphs.  For example, in the
calculation of the quantity $\partial C_{e^2} /\partial \phi$ on the
left hand side of the Nielsen identity for $Z_{e^2}$, the terms
involving $\partial m_2^2/ \partial \phi$ are no longer higher order.
The corresponding terms on the right hand side of the identity come
from contributions to $Z_{e^2}$ due to two-loop graphs and one-loop
graphs with vertices proportional to $\lambda$; both types of
contributions can be neglected for generic nonzero values of $v$.

\appendix
\reset
\section{Appendix}
\def\theequation{{A.\arabic{equation}}}

      In this appendix we show that, although individual two-loop
graphs give gauge-dependent contributions to $V^{\rm eff}_{e^6}$, their
sum is $\xi$-independent.   The first step is to identify the relevant
graphs.  All graphs with vertices proportional to $\lambda$ give
higher order contributions, and so can be omitted.  Similarly, any
graph with a loop containing only $\Phi_1$ propagators is proportional
to a power of $m_1^2$ and hence of higher order.  Finally, there is a
two-loop graph containing a ghost loop, but it is manifestly
gauge-independent.  The only nonzero graphs remaining are shown in
Fig.~3.

     It is convenient to decompose the photon propagators into
transverse and longitudinal parts, and to examine separately the
contributions from each.  Consider first the contributions involving
transverse photon propagators.  The part of graph a involving two such
propagators and the part of graph e involving one transverse photon
have already been included in the one-loop calculation by the
resummation that converted the ${\tilde m}_i^2$ to the $m_i^2$, and
hence should be omitted.  This leaves the portions of graphs a, b, and
d that involve only a single transverse photon each.  The contribution
of these to the effective potential is
\begin{eqnarray}
&&  -e^2 \int {d^4k \over (2\pi)^4} \, {d^4p \over (2\pi)^4} \,
      {g_{\mu\nu} - {k_\mu k_\nu \over k^2}
     \over (k^2 -e^2 \phi^2) \, D(p)\, [(k+p)^2 -m_1^2 ]}
    \left[ 2e^2 \phi^2 \,(\xi p^2 - e^2 v^2)\,
          \left({p_\mu p_\nu \over p^2}\right) \right.
 \nonumber \\ && \qquad   \left.
   \vphantom{\left({p_\mu p_\nu \over p^2}\right) }
     -{1\over 2}(p^2 -\xi e^2\phi^2) \, (k+2p)_\mu (k+2p)_\nu
     - 2e^2\phi (\xi \phi +v) (k+2p)_\mu p_\nu \right]
    \nonumber \\ && \qquad \, \,=
    2 e^2 \int {d^4k \over (2\pi)^4} \, {d^4p \over (2\pi)^4} \,
      {1 \over (k^2 -e^2 \phi^2) \,  [(k+p)^2 -m_1^2 ]}
    \left( 1 - {(p\cdot k)^2 \over k^2 p^2} \right) + \cdots
\end{eqnarray}
where the dots represent terms proportional to $m_2^2$.  Not only is
this result independent of $\xi$, but examination of the integrals
shows it to be in fact of order $e^8$.

   This leaves us with the terms involving only longitudinal photons.
Let us denote the corresponding contribution from graph $j$ by
$J_j$. For the first four graphs this may be written in the form
\begin{equation}
   J_j = -e^2 \int {d^4k \over (2\pi)^4} \, {d^4p \over (2\pi)^4} \,
    {1 \over [(k+p)^2 -m_1^2 ] \, D(k)\, D(p) } \, a_j(k,p) \, .
\end{equation}
Omitting terms proportional to $m_2^2$, whose effects are of higher
order, one finds that
\begin{eqnarray}
  &&  a_a = e^2 \phi^2 \, (\xi k^2 - e^2v^2)\, (\xi p^2 - e^2v^2)\,
      (k \cdot p)^2 \, k^{-2}\, p^{-2}  \nonumber \\
  &&  a_b = - {1 \over 2} \, (\xi k^2 - e^2v^2)\, ( p^2 - \xi e^2 \phi^2)\,
      (k^2 + 2 k \cdot p)^2 \, k^{-2}  \nonumber \\
  &&  a_c = {e^2 \over 2}\, (\xi \phi +v)^2\, (k^2 + 2 k \cdot p) \,
            (p^2 + 2 k \cdot p)    \nonumber \\
  &&  a_d = - 2e^2 \phi\, (\xi \phi +v)\, (\xi k^2 - e^2v^2)\,
          (k^2 + 2 k \cdot p) \, ( k \cdot p) \, k^{-2}  \, .
\end{eqnarray}
The sum of these is
\begin{eqnarray}
&&   a_a + a_b +a_c +a_d  =
   - {1\over 2} \, (p+k)^2 \,\left[ (k^2 - \xi e^2 \phi^2)\, (\xi p^2 - e^2v^2)
        + ( k \cdot p) \, (2 p^2 \xi - e^2 \phi^2 \xi ^2) \right]  \nonumber \\
&& \qquad + {\xi \over 2} \, D(p) \, [ (p+k)^2 - p^2 ]   + A(k,p) + \cdots
\end{eqnarray}
where $A(k,p)$ is an antisymmetric function of $k$ and $p$ and
the dots represent terms that are either proportional to
$m_2^2$, and thus of higher order, or else $\xi$-independent.

    When this sum is inserted back into the integral, the term
containing $A(k,p)$ vanishes because of its antisymmetry.  The
remaining terms give
\begin{eqnarray}
&&   J_a + J_b +J_c +J_d =
 {e^2 \over 2} \int {d^4k \over (2\pi)^4}
  \,  {d^4p \over (2\pi)^4} \, \left[ 1 + {m_1^2 \over  [(k+p)^2 -m_1^2 ]}
    \right]
    \, {1 \over  D(k)\, D(p) } \,
 \nonumber \\ && \qquad  \times
     \,\left[ ( k^2 -\xi e^2 \phi^2)\, (\xi p^2 - e^2v^2)
        + ( k \cdot p) \, (2 p^2 \xi - e^2 \phi^2 \xi ^2) \right]  \nonumber \\
&& \qquad - {e^2 \xi \over 2} \int {d^4k \over (2\pi)^4} \,
       {d^4p \over (2\pi)^4} \,
    { [(p+k)^2 - p^2]\over [(k+p)^2 -m_1^2 ] \, D(k) } +\cdots  \, .
\end{eqnarray}
The terms in the first integral that are proportional to $m_1^2$ are
at least $O(e^8)$ and can be omitted.  In the second integral, let us
make the change of variable $ p \rightarrow p-k$.  The resulting
integral is then clearly the product of two integrals, one of which is
proportional to $m_1^2$, and is thus also higher order.  Hence,
\begin{equation}
     J_a + J_b +J_c +J_d = {e^2\over 2} \int {d^4k \over (2\pi)^4} \,
   { (k^2 - \xi e^2 \phi^2) \over D(k) } \, \int
   {d^4p \over (2\pi)^4} \,    { (\xi p^2 - e^2v^2) \over D(p) } \,
    + \cdots  \, .
\end{equation}
This last expression is precisely equal to $-J_e$.  Hence, the
two-loop contribution to $V^{\rm eff}_{e^6}$ is $\xi$-independent, as
was claimed.

\medskip
\leftline{\large \bf Acknowledgment}
\smallskip
\noindent One of us (EJW) would like acknowledge the hospitality of
the Fermi National Accelerator Laboratory, where part of this work was
done. This work was supported in part by the U.S. Department of Energy

\newpage

\centerline{\epsffile{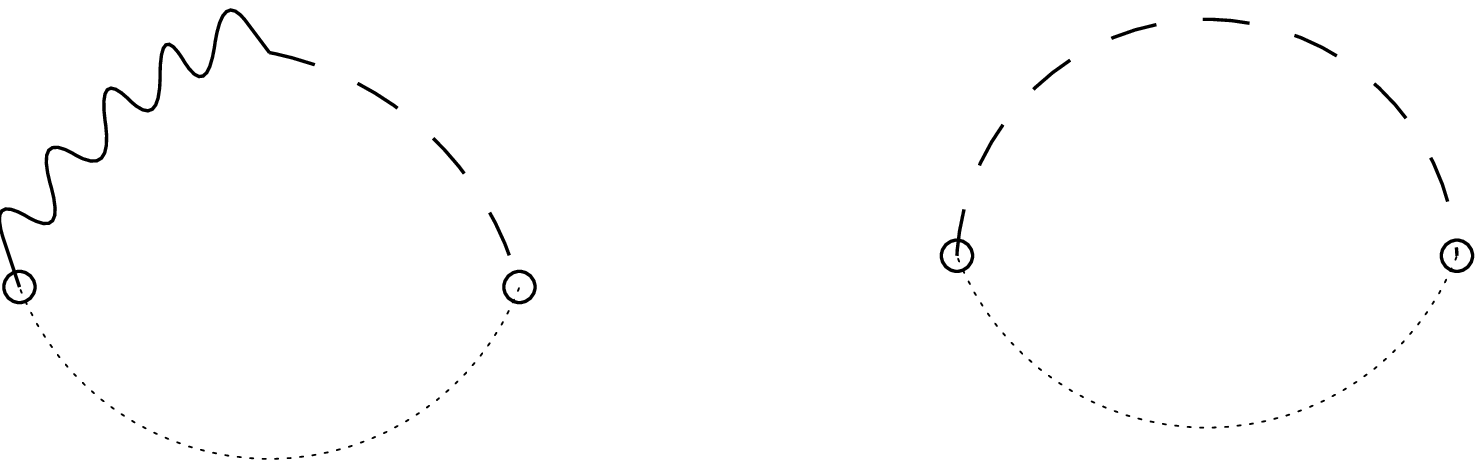}}
\bigskip \bigskip
\noindent Figure 1: The two graphs that contribute to $C_{e^2}$.  Photon,
$\Phi_2$, and ghost propagators are indicated by wiggly, long-dashed,
and short-dashed lines, respectively.

\newpage

{\epsffile{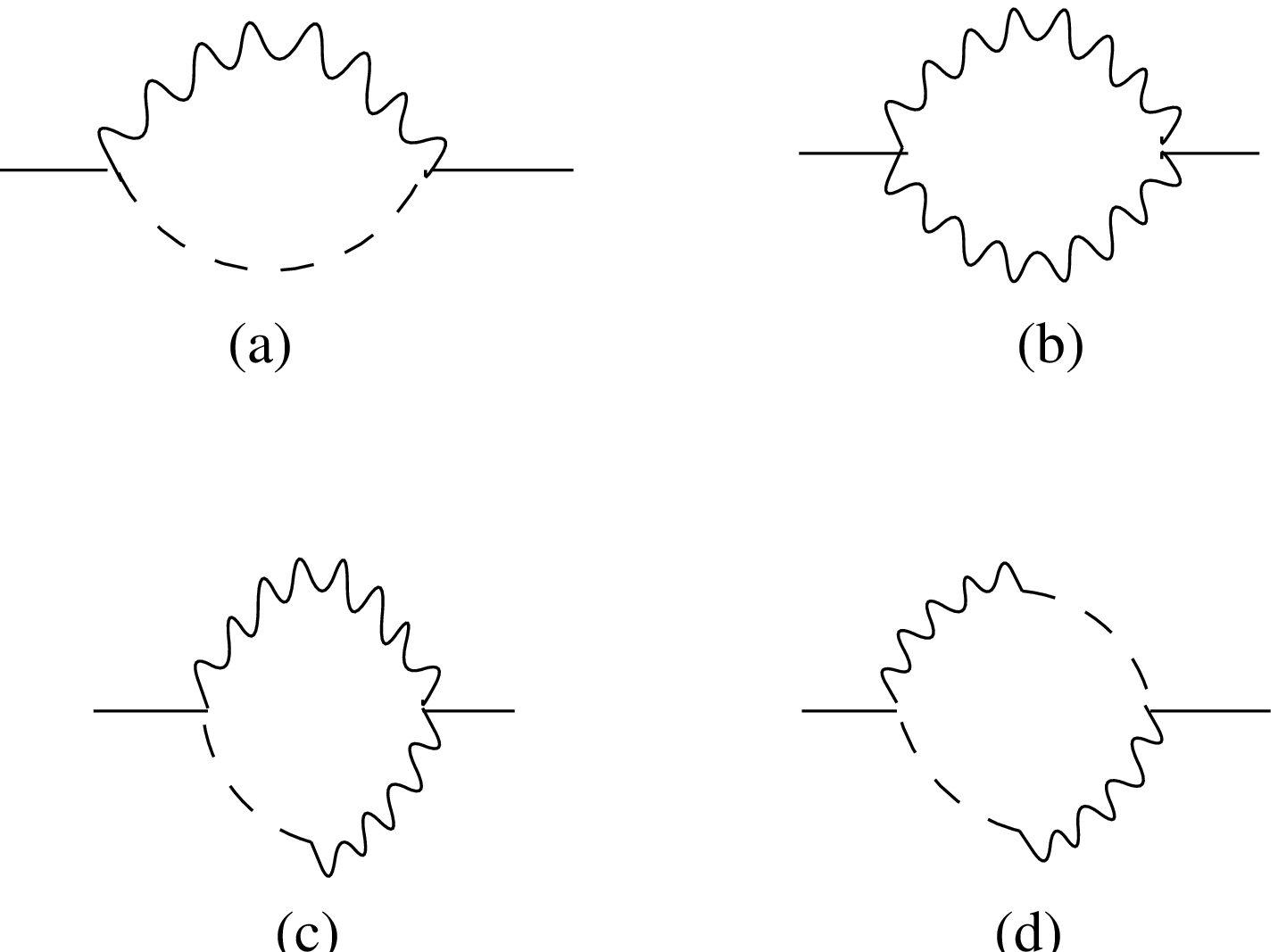}}
\bigskip\bigskip
\noindent Figure 2: The graphs that contribute to $Z_{e^2}$.  Solid lines
represent $\Phi_1$ propagators, with all other propagators shown as in
Fig.~1.
\newpage

{\epsffile{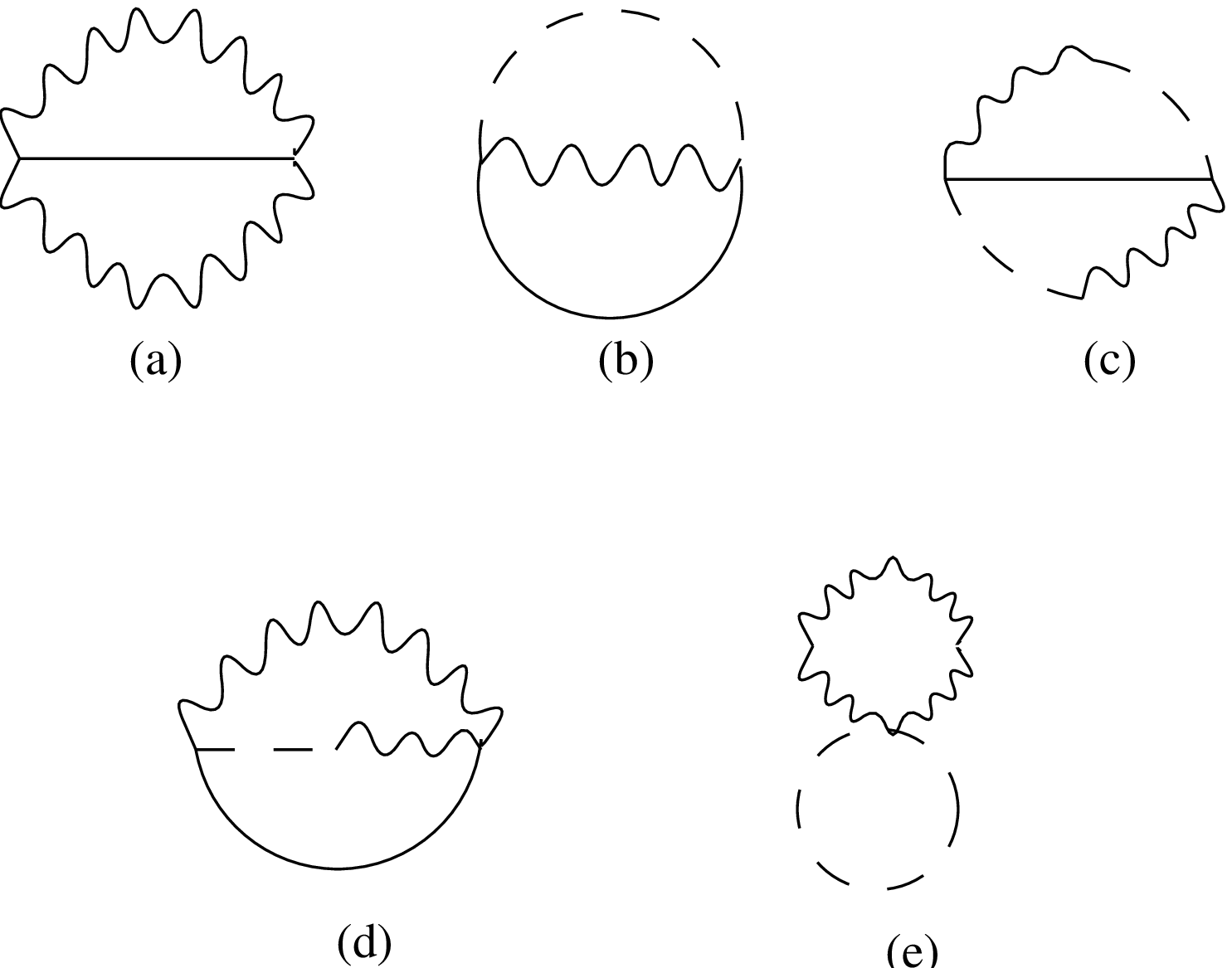}}
\bigskip\bigskip
\noindent Figure 3: The two-loop graphs that contribute to $V_{e^6}$.
\end{document}